\documentstyle[12pt]{article}

\newcommand{\beq}{\begin{equation}}          
\newcommand{\eeq}{\end{equation}}          
\newcommand{\bea}{\begin{eqnarray}}          
\newcommand{\beas}{\begin{eqnarray*}}          
\newcommand{\beau}[1]{\begin{equation} \label{#1} \begin{array}{rcl}}          
\newcommand{\eea}{\end{eqnarray}}          
\newcommand{\eeas}{\end{eqnarray*}}          
\newcommand{\eeau}{\end{array} \end{equation}}          
\newcommand{\bay}{\begin{array}}          
\newcommand{\eay}{\end{array}}

\def\L{{\cal L}}
		
\newcommand{\half}{{1 \over 2}}           
\newcommand{\Tr}{\mbox{Tr}}           
		
\newcommand{\erre}{{\hbox{{\rm I}\hspace{-.2em}\hbox{\rm R}}}}          

\newcommand{\lora}{{\longrightarrow}}

\newcommand{\lgf}{{\cal L}_{gf}}

\newcommand{\vev}[1]{\langle #1 \rangle}

\newcommand{\esp}[1]{\, e^{\textstyle #1}}

\newcommand{\di}{{\mbox{d}}}

\newcommand{\D}[1]{{\cal D} #1}          
\newcommand{\da}{{\rm d}_A}          
\newcommand{\daz}{{\mbox{d}}_{A_{0}}}          
\newcommand{\dad}{{\mbox{d}}_A^{\dagger}}          
\newcommand{\dadz}{{\mbox{d}}_{A_{0}}^{\dagger}}          
		 
\newcommand{\com}[2]{\left[ #1, #2 \right]}

\begin{document}            


\begin{titlepage}            
	   
\setlength{\textheight}{23.5cm}        
		
\begin{flushright}            
{IFUM 611/FT}  \\            
\end{flushright}            
		
\begin{center}         
{\large \bf Topological Yang-Mills cohomology in pure Yang--Mills Theory} \\    
\vspace{1cm}          
	{\bf Alberto Accardi}\footnote{E-mail:
	{\it accardi@mi.infn.it}} \\          
	{\sl Dipartimento di Fisica, Universit\`a di Milano, \\                
	Via Celoria 16 \ \ 20133 \ Milano \ \ ITALY }  \\        
	\vspace{.4cm}
	$\mbox{\bf Andrea Belli}\footnote{E-mail:
        {\it belli@mi.infn.it}} $ \\         
	{\sl Dipartimento di Fisica, Universit\`a di Milano \\            
	Via Celoria 16 \ \ 20133 \ Milano \ \ ITALY} \\          
	\vspace{.4cm}         
	{\bf Maurizio Martellini}\footnote{E-mail:           
	{\it martellini@mi.infn.it}} \\          
	{\sl Dipartimento di Fisica, Universit\`a di Milano, \\                
	I.N.F.N. \ - \  Sezione di Milano, \\            
	Via Celoria 16 \ \ 20133 \ Milano \ \ ITALY \\           
	and \\          
	Landau Network  at ``Centro Volta'', Como, ITALY}  \\         
	\vspace{.4cm}         
	{\bf Mauro Zeni}\footnote{E-mail: {\it zeni@mi.infn.it}} \\      
	{\sl Dipartimento di Fisica, Universit\`a di Milano \\            
	I.N.F.N. \ - \  Sezione di Milano, \\            
	Via Celoria 16 \ \ 20133 \ Milano \ \ ITALY}          
\end{center}         
	    
\vspace{1cm}
\begin{abstract}          
Using the first order formalism (BFYM) of the Yang-Mills theory we show that 
it displays  
an embedded topological sector corresponding to the field content of the 
Topological Yang-Mills theory (TYM). This picture arises after a proper 
redefinition of the fields of BFYM and gives a clear representation of the 
non perturbative part of the theory in terms of the topological sector. 
In this setting the calculation of the $vev$ of a YM observable is 
translated into the calculation of a corresponding (non topological) 
quantity in TYM.  
We then compare the topological observables of TYM 
with a similar set of 
observables for BFYM and discuss the possibility of describing 
topological observables in YM theory.    
\end{abstract}           
		 
\begin{flushbottom}         
\begin{footnotesize}          
\centerline{PACS: 11.10, 11.15.B}          
\centerline{Keywords: Yang-Mills, Topological BF Theory,            
Topological YM Theory}         
\end{footnotesize}          
\end{flushbottom}         
	    
\end{titlepage}          
		
\setcounter{page}{1}          
\setcounter{footnote}{0}          


\section{Introduction}         
\setcounter{equation}{0}

The search for topological quantities in field theory has been strongly 
related to the efforts to perform non perturbative calculations and in 
particular in gauge theories has been related 
to the so-called instantons calculus \cite{thooft}. It is widely believed that 
topologically non trivial configurations play a dominant role in 
phenomena like the quark confinement in QCD but the 
proper quantitative framework in which 
such a dynamics should emerge is still  missing.

Topological theories arose as
the first models in which topological quantities  
can be explicity computed \cite{witten}. The twisting procedure
derives these models from N=2 supersymmetric gauge theories but 
there is no systematic relationship with bosonic ones.  
Recently some interesting papers by Anselmi 
\cite{Anselmi-anomalies,Anselmi-TFT} tried to shed light 
on the relation between topological theories and physical ones.
In particular he suggested  that Topological Yang-Mills theory (TYM) 
could be embedded in the ordinary Yang-Mills theory (YM) to account 
for the non perturbative sectors of the theory.

More recently \cite{ccfmrtz}, along similar ideas, 
YM theory has been translated 
in the first order formalism as 
a deformation of a topological theory of BF type \cite{blau}, 
named BFYM theory. 
The full equivalence of BFYM  with the standard second order formulation 
has been proved both with 
path integral \cite{ccfmrtz} and with algebraic  methods \cite{fmstvz} 
and also the $uv$-behaviour has been checked to be the same \cite{mz}.
This formulation of YM theory has an 
enlarged symmetry and field content from which 
new  observables (inherited from the pure topological theory) 
can be defined \cite{ccgm,fmz} and is promising to start both a deeper
understanding of the topological and geometrical structure underlying
gauge theories \cite{ccr} and a new hint to disclose the long range dynamics 
of QCD. A wide review of the present status of the work on BFYM can be found 
in ref. \cite{tanzini}. 
A similar proposal to regard YM as deformation of a topological theory has been 
very recently discussed in \cite{kondo}. 

The formulation of YM as a deformation of a topological theory of BF type
strongly suggests the existence of an embedded topological sector, 
which should be related with the non perturbative, topologically non trivial 
features of the theory. 

In this paper we explicitly show how this topological sector  
arises in BFYM after a suitable field redefinition. Precisely the theory 
is decomposed in TYM plus the local quantum fluctuations and a clear 
representation of the non perturbative part of the theory in 
terms of the topological sector  is given, thus explicitly 
realizing  the conjectures of \cite{Anselmi-anomalies,Anselmi-TFT}. 
This point is discussed in section 2. 
We then discuss in section 3  the observables of BFYM. 
TYM observables can be directly compared  with a similar set in BFYM, and the 
role of local fluctuations to spoil their topological character is clearly 
displayed. More generally our framework translates the computation of the $vev$ 
of every YM observable to the computation of a related  (non topological) 
quantity in TYM. In 
this way we can formally give a set of  
sufficient conditions to be met by  a YM 
observable in order to be topological.


\section{BFYM in the self-dual gauge and relation with TYM}            
\setcounter{equation}{0} 

The classical Lagrangian of BFYM theory is given by \cite{ccfmrtz}     
\beq         
	 {\cal L}_{BFYM} = \Tr \left\{ iB\wedge F_A + g^2 
		(B - \frac{1}{\sqrt{2}g} \da\eta)         
		\wedge * (B - \frac{1}{\sqrt{2}g} \da \eta) \right\} \ ,      
 \label{BFYMaction}        
\eeq      
where $\da\equiv \di +[A, \cdot ]$ is the covariant derivative, $A$ is 
the gauge field and where antihermitian conditions have been chosen for 
the generators of the Lie algebra.  
(Wedge products will always be understood in the following). \\     
The 2-form $B$ and the 1-form $\eta$ represent the extra field 
content of the 
first order formulation of YM theory; nevertheless the physical degrees 
of freedom of the theory are not changed
and the theory is still physically equivalent to the standard 
second order formulation. Indeed an enlarged symmetry content corresponds 
to the extra degrees of freedom and requires a proper gauge fixing and ghost 
structure.
     
The corresponding BRST transformations are \cite{ccfmrtz}:         
\beau{BRSTt}         
	s \, A &=& \da c  \\         
	s \, c &=& -\half \com{c}{c}  \\         
	s \, B &=& \com{B}{c} - \da \psi + {1 \over {\sqrt{2} g}} 
		\com{F_A}{\rho} \\         
	s \, \psi &=& - \com{\psi}{c} + \da \phi  \\         
	s \, \phi &=& \com{\phi}{c} \\        
	s \, \eta &=& \com{\eta}{c}  - \sqrt{2} g \psi + \da \rho  \\        
	s \, \rho &=& - \com{\rho}{c} + \sqrt{2} g \phi \ ,        
\eeau         
and amount to both the gauge and the ``topological'' symmetries. 
The latter corresponds
to the symmetry present in the pure topological BF theory \cite{blau} 
and requires a ghost of ghost structure due to its reducible character; in 
particular the ghosts $\psi$ and $\phi$ are exactly those of pure BF theory.
Note, in comparison with the pure topological theory in which no local degrees 
of freedom are present, that a local dynamics for the BFYM theory is 
restored by the vector field $\eta$ 
and the associated ghost $\rho$. 

Moreover the field equations of (\ref{BFYMaction}) can be 
arranged \cite{ccfmrtz} as       
\beau{eqmot}        
	\dad A &=& 0  \\        
	B &=& - {i \over {2 g^2}} * F_A  \\        
	\eta &=& 0         \ ,
\eeau        
from which the moduli space ${\cal M}$ of BFYM clearly appears 
to be the same of that of YM, and from which 
the classical value of $\eta$ turns out to be zero, 
the proper on-shell value for a quantum fluctuation.

Therefore we want to interpret $\eta$ as a gluon quantum fluctuation, and 
we  interpret $A_0$ defined as $A_0 = A - \sqrt{2}g \eta$ 
as the whole gluon field        
{\it minus} its quantum fluctuations, i.e. the background connection.        
Consistently with the BRST transformations,  
$c_0 = c - \sqrt{2} g \rho$ will be regarded in a similar 
manner as the background 
gauge ghost and $\rho$ as its quantum fluctuation.  These are the first 
steps of a redefinition of the fields which will 
display the embedded topological 
sector of the 
theory. 

To quantize the theory we have also to specify the gauge fixings and 
in order to study the instanton sector of the theory we choose the          
self-dual gauge-fixing $B^-=0$ for the topological symmetry and a covariant     
gauge-fixing for the remaining symmetries; explicitly, the conditions we 
choose are:         
\beau{gfc}         
	\di_{A-\sqrt{2}g\eta}^\dagger A &=& 0  \\         
	B^- &=& 0  \\         
	\di_{A-\sqrt{2}g\eta}^\dagger \psi &=& 0  \\         
	\di_{A-\sqrt{2}g\eta}^\dagger \eta &=& 0 \ , \\          
\eeau        
properly expressed in terms of the background  connection $A_0$. 
           
Moreover, we implement the conditions (\ref{gfc}) 
introducing the following BRST doublets        
\beau{bd}        
	s \, \bar{c} &=& h_A  \\         
	s \, h_A &=& 0  \\         
	s \, \bar{\chi} &=& h_B  \\        
	s \, h_B &=& 0 \\        
	s \, \bar{\phi} &=& h_\psi \\        
	s \, h_\psi &=& 0  \\        
	s \, \bar{\rho} &=& h_\eta  \\        
	s \, h_\eta &=& 0 \ ,        
\eeau
(in particular $\bar{\chi}$ and $h_B$ are anti--self--dual 2--forms)         
and the Landau gauge--fixing Lagrangian:        
\beq        
	\lgf = \Tr \left\{ s \left( \bar{c}* \di_{A_0}^\dagger A        
		+ \bar{\chi} * B^-        
		+ \bar{\phi} * \di_{A_0}^\dagger \psi        
		+ \bar{\rho} * \di_{A_0}^\dagger \eta  \right) \right\} \ .    
\eeq

We now complete the change of variables in 
order to study the formal relation between the BFYM and        
TYM. This change will isolate 
a subset of the new fields having  the same BRST transformations        
as the fields appearing in TYM theory. These fields will be interpreted as     
background fields for the quantum fluctuations of the BFYM theory. 

The full change of variables is the following \footnote{This change of        
variables was suggested in \cite{ccfmrtz}.}:        
\beau{cov}        
	A_0 &=& A - \sqrt{2} g \eta \\        
	c_0 &=& c - \sqrt{2} g \rho \\        
	\phi_0 &=& - \phi + \half \com{\rho}{\rho} \\        
	\psi_0 &=& \psi - \com{\eta}{\rho} \\        
	\eta' &=& \eta \\         
	\rho' &=& \rho \ .        
\eeau 
The first two equations correspond to the previous expansion of the fields 
$A$ and $c$ in a background and a quantum fluctuation part. Note that also 
the other tranformations are entirely given in terms of the fluctuations 
$\eta$ and $\rho$. Moreover the Jacobian of the transformation being 1, 
the functional measure is\footnote{In the following we        
will omit the primes.}        
\beq        
	\D{A} \D{c} \D{\psi} \D{\phi} \D{\eta} \D{\rho}        
		=  \D{A_0} \D{c_0} \D{\psi_0} \D{\phi_0} \D{\eta'}       
		   \D{\rho'} \ .        
\eeq        
The BRST algebra in terms of the new variables becomes        
\beau{BRSTzero}         
	s A_0 &=& \daz c_0 + 2 g^2 \psi_0 \\         
	s c_0 &=& - \half \com{c_0}{c_0} + 2 g^2 \phi_0 \\         
	s \psi_0 &=& - \com{\psi_0}{c_0} - \daz \phi_0  \\         
	s \phi_0 &=& \com{\phi_0}{c_0} \\        
	&& \\        
	s B &=& \com{B}{c_0} - \daz \psi_0 - \sqrt{2} g \com{\eta}{\psi_0} 
		+ {1 \over {\sqrt{2} g}} \com{F_{A_{0}}}{\rho} + 
		\com{\eta}{\daz \rho} + \sqrt{2} g \com{B}{\rho} \\         
	s \eta &=& \com{\eta}{c_0}  - \sqrt{2} g \psi_0 + \daz \rho
		+ \sqrt{2} g \com{\eta}{\rho} \\        
	s \rho &=& - \com{\rho}{c_0} - \sqrt{2} g \phi_0 - 
		{g \over {\sqrt{2}}} \com{\rho}{\rho} \ .        
\eeau         
This is the key result of the present paper: 
the fields ($A_0$, $c_0$, $\psi_0$, $\phi_0$) correspond to the 
field content of TYM and their BRST transformations are exactly  
those of TYM \cite{witten}
(modulo the rescaling $2 g^2 \psi_0 \lora \psi_0$, 
$2 g^2 \phi_0 \lora \phi_0$). These fields clearly display an 
embedded topological sector  in the YM theory; moreover a direct 
comparison among the observables of YM and TYM is now available and 
will be discussed in the next section.

Following the previous interpretation we separate the fields into two 
classes, the ``background'' fields $\varphi_0$ 
(including those of TYM) and the local fluctuations 
$\varphi_{q}$, and express the action as the sum of a background action 
$S_0 = S_0[\varphi_0]$     
containing all the terms depending only on $\varphi_0$ and a fluctuation        
action $S_{q}=S_q[\varphi_q;\varphi_0]$ made of the remaining terms.        
With this decomposition the functional integral becomes:        
\beq        
	Z = \int \D{\varphi_0} \esp{-S_0[\varphi_0]} \int \D{\varphi_q}     
		\esp{-S_q[\varphi_q;\varphi_0]}  \ .
\eeq        
Note that the two actions aren't decoupled. 
Following the physical interpretation, in an explicit computation  we could 
make a saddle point 
expansion of the second integral over the background $\varphi_0$ 
and then make the integration on $\D{\varphi_0}$. The idea is that 
this  
separation of fluctuations from backgrounds could lead 
to a deeper understanding of the  role of the topological sector 
(i.e. the content of TYM)  
in the non perturbative calculations  of YM theory.

Now we turn to the $B$-field: how have we to treat it?        
We should decompose it into a background 
and a fluctuation part transverse with 
respect to the former, 
\beq        
	B = B_0 \oplus B_{q}\ ,
 \label{Bdec}        
\eeq        
such that the functional measure over $B$ factorize:        
\beq        
	\D{B} = \D{B_0} \D{B_q} \ .        
\eeq        
A natural choice in the self-dual gauge-fixing is:        
\beq        
	B_0 = B^+  \ \ \ \ \ \  B_q = B^- \ .         
 \label{Bdecomp}        
\eeq         
The decomposition (\ref{Bdecomp}) has the advantage that the gauge-fixing       
conditions set $B_q = 0$. 
Other decompositions are available:  for example we could assign        
$s \, B_0 = \com{B_0}{c_0} + \daz \psi_0$ and obtain the algebra of the 
topological BF     
theory with cosmological term. Such a decomposition should be useful in a      
covariant gauge, but is more difficult to implement in the functional        
integral.        
 
We then choose (\ref{Bdecomp}) and redefine the (\ref{bd}) as:        
\beau{bd1}        
	\bar{c}_0 &=& \bar{c}  \\         
	h_{A_0} &=& h_A  \\         
	\bar{\phi}_0 &=& \bar{\phi}  \\        
	h_{\psi_0} &=& h_\psi  \\        
	\bar{\chi}' &=& \bar{\chi} \\        
	h'_B &=& h_B \\        
	\bar{\rho}' &=& \bar{\rho} + \sqrt{2} g \bar{c}  \\        
	h'_\eta &=& h_\eta + \sqrt{2} g h_A  \ ,        
\eeau        
in such a way that they remain BRST doublets. Then we set         
\beas        
	\varphi_0 &=& \left( A_0, c_0, \psi_0, \phi_0, B^+, \bar{c}_0,        
		h_{A_0}, \bar{\phi}_0, h_{\psi_0}, \bar{\chi}'  \right) \\   
	\varphi_q &=& \left( \eta, \rho, B^-, \bar{\rho}', h'_\eta, 
		h'_B \right) \ ,
\eeas        
and discuss first the classical action and then the gauge-fixing one.         

By substituting (\ref{cov}) and (\ref{Bdec},\ref{Bdecomp}) into the 
classical Lagrangian (\ref{BFYMaction}) and imposing the $B^- = 0$ 
condition we get    
\beq
	{\cal L}_{BFYM} \equiv {\cal L}_0 + {\cal L}_q 
\eeq
with
\bea
	{\cal L}_0 &=& i B^+ F_{A_{0}} + g^2 B^+ * B^+ \\       
	{\cal L}_q &=& i B^+ \left( \sqrt{2} g \daz \eta + g^2 \com{\eta}{\eta} 
		\right) - B^+ * \left( \sqrt{2} g \daz \eta + 2 g^2 
		\com{\eta}{\eta} \right) + \half \daz \eta * \daz \eta + 
		\nonumber \\
	&& + \sqrt{2} g \daz \eta * \com{\eta}{\eta} + g^2 \com{\eta}{\eta} * 
		\com{\eta}{\eta} \ ,        
\eea        
where ${\cal L}_0$ is the classical Baulieu-Singer TYM action written in the    
first order formalism \cite{witten} 
and ${\cal L}_{q}$ is the action on the quantum fluctuations and corresponds 
to the classical YM action expanded around a      
background connection $A_0$ with quantum fluctuation $\sqrt{2} g\eta$.         
The gauge-fixing Lagrangian becomes instead:        
\beq
	\L^{gf} \equiv \L^{gf}_0 + \L^{gf}_q
	\label{Lgfz}
\eeq
with
\bea        
	\L_0^{gf} &=& \Tr \left\{ h_{A_{0}} * \dadz A_0 + 
		h_{\psi_{0}} * \dadz \psi_0 - \bar{c}_0 * \dadz 
		\daz c_0 - 2 g^2 \bar{c}_0 * \dadz \psi_0 + \right.
		\nonumber \\
	&& \hspace{0.7cm} - \bar{c}_0 \com{\daz c_0}{* A_0} - 2 g^2 
		\bar{c}_0 \com{\psi_0}{* A_0} + \bar{\phi}_0 * \dadz 
		\daz \phi_0 - 2 g^2 \bar{\phi}_0 \com{\psi_0}{* \psi_0} + 
		\nonumber \\
	&& \hspace{0.7cm} \left. + \bar{\phi}_0 * \com{\dadz \psi_0}{c_0} 
		+ \bar{\chi}' * \left( 	\daz \psi_0 \right)^- \right\} \ , \\
	\L_q^{gf} &=& \Tr \left\{ h_{\psi_{0}} * \com{\dadz \eta}{\rho} 
		+ h'_\eta * \dadz \eta + \sqrt{2} g 
		\bar{\phi}_0 * \com{\dadz \psi_0}{\rho} + \right.
		\nonumber \\
	&& \hspace{0.7cm} + \sqrt{2} g \bar{\phi}_0 *  
		\com{\dadz \eta}{\phi_0} - \half \bar{\phi}_0 * \dadz \daz 
		\com{\rho}{\rho} - \bar{\phi}_0 * 
		\com{c_0}{\com{\dadz \eta}{\rho}} + \nonumber \\
	&& \hspace{0.7cm} - 2 g^2 \bar{\phi}_0  
		\com{\psi_0}{* \com{\eta}{\rho}} - \bar{\rho}' * 
		\com{\dadz \eta}{c_0} - 2 g^2 \bar{\rho}'  
		\com{\psi_0}{* \eta} - \bar{\rho}' * \dadz \daz \rho + 
		\nonumber \\
	&& \hspace{0.7cm} + \sqrt{2} g \bar{\rho}' * \dadz \psi_0 - \sqrt{2} g 
		\bar{\rho}' * \com{\dadz \eta}{\rho} + 
		\nonumber \\
	&& \hspace{0.7cm} \left. 
                + \sqrt{2} g \bar{\chi}' * \com{\eta}{\psi_0}^- - 
		{1 \over {\sqrt{2} g}} \bar{\chi}' * \com{F^-_{A_{0}}}{\rho} 
		 - \bar{\chi}' * \com{\eta}{\daz \rho }^- \right\} \ .        
\eea         
The Lagrangian ${\cal L}_0^{gf}$ is quite equal to the gauge-fixing        
Lagrangian of TYM quantized in the gauge         
$\dadz A_0 =0$, $ F_{A_0}^- =0$ and $ \dadz \psi_0 = 0$. 
The only difference is that the 
terms $- \bar{\chi} * \com{F^-_{A_{0}}}{c_0}$ and $h_{F} * F^-_{A_{0}}$ are 
lacking and we have $- \bar{\chi}' * \com{B^-}{c_0}$ and $h'_B * B^-$ instead, 
all these terms  vanishing in the Landau gauge. 
Thus we can read ${\cal L}_0^{gf}$ as the TYM self--dual gauge--fixing 
Lagrangian in the first order formalism and,  
after the $h'_B$ and $B^-$ integrations,
\beq        
	{\cal L}_0^{gf} \equiv {\cal L}_{TYM}^{gf}        \quad .
\eeq        

In summary the partition function 
is:        
\bea        
	Z \!\! &=& \!\! \int \D{A_0} \D{B^+} \D{c_0} \D{\psi_0} \D{\phi_0} 
		\D{\bar{c}_0} \D{\bar{\phi}_0} \D{h_{A_{0}}} \D{h_{\psi_{0}}}   
		 \esp{- \int \di^4x (iB^+F_{A_0}+g^2 B^+*B^+) + S_0^{gf}}
		\cdot \nonumber \\        
	&& \cdot \int \D{\eta} \D{\rho} \D{\bar{\rho}'} \D{\bar{\chi}_0}
		\D{h'_\eta} \esp{- S_q[\varphi_q;\varphi_0] }  \ ,
       \label{last}       
\eea        
where $S_0^{gf} = \int \di^4x \L_0^{gf}$ and 
$S_q = \int \di^4x \left( \L_q + \L^{gf}_q \right)$. Moreover
if we would like to compute a YM observable, i.e. an observable not        
containing $B^+$, or an observable linear in $B$ then 
we could integrate over $B^+$ in  $ \L_0 + \L^{gf}_0$ 
and obtain {\it exactly} the standard second order TYM action plus 
fluctuations. 
Therefore we see that the YM theory has been recast in the 
theory given by  eq.(\ref{last}) where the nested integration corresponds
to the contribution of local fluctuations on a background 
described by a topological theory. 
In the next section we discuss the 
observables of BFYM theory in this framework.

\section{Observables} 
\setcounter{equation}{0} 
	    
\subsection{YM observables}    
    
We define a {\it YM functional} a functional of the form    
\beq    
	{\cal O} = {\cal O} [A,c,\bar{c}]     
		= {\cal O} [\eta,\rho; A_0,c_0,\bar{c}_0]    
\eeq    
i.e. a functional constructed out of the fields that are naively identified     
with the YM ones (all the other fields are added by gaussian integration     
to the YM action to obtain the action of BFYM).  \\    
Then we consider the {\it YM amplitude} $\vev{\cal O}$ and recall that for 
$B$--independent amplitudes we can integrate out $B^+$ and 
obtain:    
\bea    
	\vev{\cal O} &=& \int \D{\varphi_0} 
		\esp{-S_{TYM}[\varphi_0]}    
		\int \D{\varphi_q}     
		\esp{-S_q[{\varphi_q};{\varphi_0}]}    
		{\cal O} [\eta,\rho; A_0,c_0,\bar{c}_0]       
		\nonumber  \\    
	&\equiv & \int \D{\varphi_0} 
		\esp{-S_{TYM}[{\varphi_0}]}    
		{\cal A}_{\cal O}[{\varphi_0}] \ ,      
		\label{YMampl} 
\eea    
where we can think the amplitude ${\cal A}_{\cal O}$ computed for instance     
perturbatively. \\    
Note that we have reduced the evaluation of an YM amplitude 
in BFYM theory to the   
computation of an amplitude in TYM theory; as it is well known, if the 
amplitude were  topological the     
only contribution to it would come from the moduli space (or in other     
words the semiclassical approximation would be exact). 
In our case this will be not true  because the amplitude 
${\cal A}_{\cal O}$ will not be in general an observable in the TYM sense, 
i.e. it will not be topological.

The ordinary perturbation theory corresponds to the the $k=0$ sector of 
the BFYM in which case the zero instantons moduli     
space is ${\cal M}_0 = \{ A_0 = 0 \}$. 
In general only the terms which saturate the ghost anomaly contribute to 
the amplitude and in this case the computation of  (\ref{YMampl}) 
is performed 
expanding  $\esp{-S_q[\varphi_q;\varphi_0]}$     
and $\cal{O}$ in powers of the background ghosts and setting $A_0 = 0$; since   
${\cal M}_0$ has no ghost anomaly we can retain only the zero order terms. 
When considering YM amplitudes in the higher instantons sectors  
we are similarly led to a computation of a certain amplitude in TYM. 
In this case we have however to saturate the fermionic     
anomaly of the moduli space ${\cal M}_k$.  
We can do it in two ways.  
\begin{itemize} 
  \item We can expand ${\cal A}_{\cal O}[\varphi_0]$ 
in powers of the background  
ghosts and retain only the terms of ghost number equal to the moduli space  
dimension (the other terms give a zero contribution when evaluated in TYM).  
  
\item We can consider a YM amplitude with the     
insertion of an observable $\cal G$ of background ghost number equal to     
${\rm dim} {\cal M}_k$. Then     
\beq    
	\vev{{\cal G} {\cal O}} = \int \D{\varphi_0} {\cal G}     
		{\cal A}_{\cal O}^{YM}[A_0] \esp{-S_{TYM}[{\varphi_0}]}    
\eeq    
where ${\cal A}_{\cal O}^{YM}[A_0]$ 
is the vacuum expectation value of the YM     
functional ${\cal O}[A,c,\bar{c}]$ in YM theory expanded over the background  
connection $A_0$. Therefore in this case we can interpret $\D{\varphi_0}  
{\cal G}$ as the measure over the instanton moduli space of the YM theory.    
\end{itemize} 
    
\subsection{Topological observables}   

Having translated the computation of correlators in YM theory into 
that of related 
quantities in TYM,   the interesting question  arises whether  topological 
amplitudes in YM theory exist. Moreover in our framework we can directly 
consider the set of topological observables of TYM which, 
after a proper identification of the fields $\varphi_0$, 
is given by \cite{witten}
\beau{TYM observables}        
	T^0_4 &=& \Tr \left(  \frac{1}{8 g^4} F_{A_0}F_{A_0} \right)  \\
        T^1_3 &=& \Tr \left( - \frac{1}{2 g^2} F_{A_0}\psi_0 \right)  \\        
	T^2_2 &=& \Tr \left( - \frac{1}{2 g^2} F_{A_0}\phi_0   
		+ {1 \over 2} \psi_0\psi_0 \right)     \\        
	T^3_1 &=& \Tr \left( \psi_0 \phi_0 \right)  \\        
	T^4_0 &=& \Tr \left( \half \phi_0 \phi_0 \right) \ .        
\eeau        
They satisfy the following descent equations:        
\beau{desceqtym}        
	s T^0_4 + \di T^1_3 &=& 0  \\        	
        s T^1_3 + \di T^2_2 &=& 0  \\        
	s T^2_2 + \di T^3_1 &=& 0  \\        
	s T^3_1 + \di T^4_0 &=& 0  \\        
	s T^4_0 &=& 0 \ . \\        
\eeau 
Apart from some field rescaling, they are the same that have been studied in  
\cite{Anselmi-anomalies,Anselmi-TFT}, where it is shown that in TYM 
they give multi-link invariants of submanifolds of $\erre^4$. 
 
A similar set of observables can be found also in BFYM. They are   
\beau{BFYMobs}        
	K^0_4 &=& \Tr \left( {1 \over {\sqrt{2} g}} \left( \da B \right) \eta 
		+ \half B B - {1 \over {2 g^2}} F_A \eta \eta \right)  \\      
	K^1_3 &=& \Tr \left( B \psi - {1 \over {\sqrt{2} g}} \left( \da B 
		\right) \rho \right)  \\        
	K^2_2 &=& \Tr \left( - B \phi + \half \psi \psi - {1 \over {2 g^2}} 
		F_A \rho \rho \right)  \\        
	K^3_1 &=& \Tr \left( - \psi \phi \right)  \\        
	K^4_0 &=& \Tr \left( \half \phi \phi \right) \ .        
\eeau        
These observables satisfy the following descent equations:        
\beau{desceq}        
	s K^0_4 + \di K^1_3 &=& 0  \\        
	s K^1_3 + \di K^2_2 &=& 0  \\        
	s K^2_2 + \di K^3_1 &=& 0  \\        
	s K^3_1 + \di K^4_0 &=& 0  \\        
	s K^4_0 &=& 0 \ . \\        
\eeau         
	    
The choice of the (\ref{BFYMobs}) is dictated by its formal ressemblance        
with those of TYM. 
Indeed, if we take $\eta=\rho=0$ (i.e. vanishing fluctuations) 
and        
substitute $B$ with $F_A$, we     
obtain up to moltiplicative constants  exactly the TYM observables.         

These observables are related in the following way: 
\bea 
	K^1_3 &=& - T^1_3 + s \ \Tr \left\{ \frac{1}{2\sqrt{2}g^3} F_A \eta 
		- {1 \over {\sqrt{2}g}} B \eta + {1 \over {3 \sqrt{2} g}}
		\eta \eta \eta \right\} + \nonumber \\
	&& + \di \ \Tr \left\{ - \frac{1}{2\sqrt{2}g^3} F_A \rho + 
		{1 \over {\sqrt{2} g}} B \rho - {1 \over {\sqrt{2}g}} \psi 
		\eta  -  {1 \over {\sqrt{2}g}}   
		\eta \eta \rho  \right\} \\
	K^2_2 &=& T^2_2 - s \ \Tr \left\{ \frac{1}{2\sqrt{2}g^3} F_A \rho +
		{1 \over {\sqrt{2} g}} B \rho - {1 \over {2 g^2}} \left( \da
		\eta \right) \rho + \frac{1}{\sqrt{2}g} \eta \eta \rho 
		\right\} + \nonumber \\
	&& - \di \ \Tr \left\{ \frac{1}{\sqrt{2}g} 
                \eta \rho \rho  
		\right\} \\
	K^3_1 &=& T^3_1 + s \ \Tr \left\{ \frac{1}{\sqrt{2}g} \eta \rho \rho 
		\right\} - \di \ \Tr \left\{ \frac{1}{3\sqrt{2}g} \rho \rho 
		\rho \right\} \\ 
	K^4_0 &=& T^4_0 + s \ \Tr \left\{ \frac{1}{3\sqrt{2}g} \rho \rho \rho
		\right\} \ . 
\eea 
We see in general that the difference between the sets is given in terms 
of the dressing due to the quantum fluctuations $\eta$ and $\rho$. 
Note also that the  $K$'s  are equivalent to the  $T$'s  
modulo $d$--exact and $s$--exact terms, hence the cohomology of YM includes  
that of TYM.

In order to study the topological properties of an observable $T$ we have 
to consider the correlator 
\bea    
	\vev{T} &=& \int \D{\varphi_0} \esp{-S_{TYM}[\varphi_0]}    
		\int \D{\eta} \D{\rho} \D{\bar{\rho}}     
		\esp{-S_q[\eta,\rho,\bar{\rho};\varphi_0]}    
		{T} [\eta,\rho,\bar{\rho},A_0,c_0,\bar{c}_0]      
		\nonumber  \\    
	&\equiv& \int \D{\varphi_0} \esp{-S_{TYM}[\varphi_0]}    
		{\cal T}_T[\varphi_0]      
		\label{topo} \ .     
\eea    
The observable $T$ in YM corresponds to ${\cal T}_T$ in TYM. Then we can give 
implicitly
on ${\cal T}_T$ the sufficient conditions for $T$ to be topological. In TYM 
topological quantities must be closed under TYM-BRST transformations and 
their metric
dependence must be TYM-BRST exact. These are precisely the requirements on 
${\cal T}_T$; when 
fullfilled, the corresponding $T$ should be a topological operator in YM. 
The TYM-BRST transformations in our case are precisely the 
nihilpotent subalgebra of (\ref{BRSTzero}):
\bea
	s_0 A_0 &=& \daz c_0 + 2 g^2 \psi_0 \\         
	s_0 c_0 &=& -\half [c_0,c_0] + 2g^2 \phi_0 \\         
	s_0 \psi_0 &=& - [\psi_0,c_0] - \daz \phi_0  \\         
	s_0 \phi_0 &=& [\phi_0,c_0]  \\        
	s_0 \, \mbox{other} &=& 0 \ .
\eea

This is the only nihilpotent subalgebra which includes the topological 
ghosts. We can then formally rewrite the ``topological'' conditions as
\bea 
	& &sT=0 \\
	& &s_0{\cal T}_T=0\\
	 & &\delta_g {\cal T}_T= s_0 f\ ,
\eea  
with arbitrary $f$, where $\delta_g$ is the derivative with respect to the 
metric. It would be very important to rewrite these conditions directly 
in terms 
of $T$ to work out the possible topological quantities. Clearly also the pure 
observables of TYM in this framework are no more topological, owing to 
the local contribution of the action on the quantum fluctuations.

 
\section{Conclusions}            
\setcounter{equation}{0} 
	   
In this letter we have displayed the topological sector embedded in 
Yang--Mills theory. Starting from the first--order (BFYM) formulation and 
performing a suitable field redefinition  we find that a subset of the fields 
of BFYM represents the field content 
of Topological Yang--Mills theory; moreover the  BRST algebra of YM 
theory on these fields reduces to that of TYM. The whole theory can then be 
split in the integration  over these ``topological'' degrees of freedom and 
over the remaining fields, which are correctly identified as the local 
quantum fluctuations that restore a local dynamics on the topological theory. 
The non perturbative sector of the theory should be naturally related with this 
topological sector; in any case we give an explicit realization of the 
relationship of topological theories with physical ones in the bosonic case. 
In this framework the calculation of correlators in YM theory is translated 
in the calculation of non topological correlators in TYM. We also 
explicitly compare the observables of TYM with a similar set in BFYM theory; 
the two sets are cohomologically equivalent and differ only by the dressing 
due to the 
local quantum fluctuations, thus showing the inclusion of the cohomology of 
TYM in that of YM theory. In this framework we have also discussed the 
possibility to have topological correlators in YM 
and found a set of sufficient 
criteria to identify them, provided that any solution exists. 

The rich structure exhibited by 
the first order formulation of Yang-Mills theory 
is currently investigated  in several respects \cite{inprogress}. 
First of all, we think that our decomposition of YM into TYM theory 
plus fluctuations could 
improve the old problem of finding a well defined  bosonic measure over the 
instanton moduli space. In this case the ``topological'' symmetry and the 
related gauge fixing procedure should correspond to a symmetry in the istanton 
moduli space requiring a proper treatment in order to give a finite 
integration volume. 

A second very interesting point is to clarify the relationship with the 
Seiberg-Witten analysis of N=2 susy YM  theories \cite{sw}. 
The twisting operation links 
N=2 susy YM to TYM theory 
and should therefore have some counterpart also in the 
YM case, the ghost of ghost structure present in BFYM providing the 
field content corresponding to that of susy multiplets. Such a relationship is 
clearly relevant to the confinement issues in QCD.

Finally, new nonlocal observables directly inherited from the topological BF 
theory can be considered in BFYM theory \cite{ccgm,fmz,ccr} and generalize 
the Wilson loop operator and the linking observables of knot theory. Again 
their analysis should be relevant for the long range features of the theory 
and indeed an area law behaviour for the $vev$ of the Wilson loop has 
been derived in this framework \cite{fmz,ccr}. We believe that these 
non local observables constitute the bridge which should connect the 
microscopical description of QCD to the language of long range hadron physics.
	   
\vskip 1.5cm            
\leftline{\bf\large Acknowledgments}            
\vskip .5cm            
		
We acknowledge useful discussions with A. Tanzini about BFYM operators and 
their descent equations. We acknowledge also useful discussions with 
A.S.~Cattaneo and P.~Cotta-Ramusino. M.M. acknowledges F.~Fucito and 
S.P.~Sorella for discussions and the kind hospitality at the 
Caxambu Meeting 97. 
M.Z. has been partially supported by MURST and by
TMR programme  ERB-4061-PL-95-0789.
		

\end{document}